


\overfullrule0pt
\input phyzzx
\physrev
\def\ls#1{\ifmath{_{\lower1.5pt\hbox{$\scriptstyle #1$}}}}
\def\ifmath#1{\relax\ifmmode #1\else $#1$\fi}
\def\DESY{\address{Deutsches Elektronen-Synchrotron, Notkestra\ss e 85,
22603 Hamburg, Germany}}

\def\calo{{\cal O}}
\def\abs#1{\left|#1\right|}
\def\d{{\rm d}}
\def\diag{{\rm diag}}
\def\det{{\rm det}}
\def\gev{{\rm  GeV }}
\def\tev{{\rm  TeV }}
\def\hc{{\rm  h.c.}}


\def\hl{{h^0}}
\def\ha{{A^0}}
\def\hh{{H^0}}
\def\hpm{{H^{\pm}}}


\def\mhl{m_{h^0}}
\def\mhh{m_{H^0}}
\def\mh{m_h}
\def\mhpm{m_{H^{\pm}}}
\def\mha{m_{A^0}}
\def\mw{m_{\rm w}}
\def\mz{m_{\rm z}}
\def\mgut{M_{\rm GUT}}
\def\msusy{M_{\rm SUSY}}


\def\sw  {s_{\rm w}}
\def\cw  {c_{\rm w}}
\def\cotb{\cot\beta}
\def\tanb{\tan\beta}
\def\sinb{\sin\beta}
\def\cosb{\cos\beta}


\def\sevenhalf{\ifmath{{\textstyle{7 \over 2}}}}
\def\eninths{\ifmath{{\textstyle{8 \over 9}}}}
\def\ninth{\ifmath{{\textstyle{1 \over 9}}}}
\def\fourninths{\ifmath{{\textstyle{4 \over 9}}}}
\def\sixteenninths{\ifmath{{\textstyle{16 \over 9}}}}
\def\fivefourths{\ifmath{{\textstyle{5\over 4}}}}
\def\fivethirds{\ifmath{{\textstyle{5\over 3}}}}
\def\half{\ifmath{{\textstyle{1 \over 2}}}}
\def\fivehalf{\ifmath{{\textstyle{5 \over 2}}}}
\def\fivetwelfth{\ifmath{{\textstyle{5 \over 12}}}}
\def\thirteenthirds{\ifmath{{\textstyle{13\over 3}}}}
\def\fourthirds{\ifmath{{\textstyle{4 \over 3}}}}
\def\third{\ifmath{{\textstyle{1 \over 3}}}}
\def\twothirds{\ifmath{{\textstyle{2\over 3}}}}
\def\fourth{\ifmath{{\textstyle{1\over 4}}}}
\def\threefourth{\ifmath{{\textstyle{3\over 4}}}}
\def\fifteenfourth{\ifmath{{\textstyle{15\over 4}}}}
\def\fifth{\ifmath{{\textstyle{1 \over 5}}}}
\def\twelfth{\ifmath{{\textstyle{1 \over 12}}}}
\def\eighth{\ifmath{{\textstyle{1 \over 8}}}}
\def\threeighth{\ifmath{{\textstyle{3 \over 8}}}}
\def\threesixtn{\ifmath{{\textstyle{3 \over16}}}}
\def\eightthirds{\ifmath{{\textstyle{8 \over 3}}}}
\def\thirtytwoninths{\ifmath{{\textstyle{32\over 9}}}}
\def\eightthirds{\ifmath{{\textstyle{8 \over 3}}}}
\def\twentythirds{\ifmath{{\textstyle{20\over 3}}}}
\def\eightninths{\ifmath{{\textstyle{8 \over 9}}}}
\def\twentyninths{\ifmath{{\textstyle{20\over 9}}}}
\def\ninehalf{\ifmath{{\textstyle{9 \over 2}}}}
\def\seventeentwelfth{\ifmath{{\textstyle{17 \over12}}}}
\def\ninefourth{\ifmath{{\textstyle{9 \over 4}}}}
\def\threehalf{\ifmath{{\textstyle{3 \over 2}}}}
\def\sixth{\ifmath{{\textstyle{1 \over 6}}}}
\def\sevensixths{\ifmath{{\textstyle{7 \over 6}}}}
\def\fivesixths{\ifmath{{\textstyle{5 \over 6}}}}
\def\seventhirds{\ifmath{{\textstyle{7 \over 3}}}}
\def\sevenhalf{\ifmath{{\textstyle{7 \over 2}}}}

\def\alphas{\alpha_{\rm s}}
\def\alphaem{\alpha_{\rm em}}
\def\tildev{\widetilde V}
\def\dmh{\left(\Delta\mhl^2\right)}
\def\mweak{m_{\rm weak}}
\def\lnbar{{\overline{\rm ln}\,}}
\date={}
\titlepage
\hoffset=-.65cm
\voffset=-.4cm
\hsize=17cm
\vsize=23.5cm
\line{DESY 93-162 and TTP93-35\hfill ISSN 0418-9833}
\line{hep-ph/9401219}
\line{November 1993\hfill}
\vskip1.cm

\title{\bf
Two--Loop Radiative Corrections to the Lightest
Higgs Boson Mass in the Minimal Supersymmetric Model}
\author{Ralf Hempfling$^1$ and Andr\'e H. Hoang$^2$}

\vskip .1in
\address{$^1$Deutsches Elektronen-Synchrotron, Notkestra\ss e 85,
D-22603 Hamburg, Germany}
\smallskip
\address{$^2$Institut f\"ur Theoretische Teilchenphysik, Universit\"at
Karlsruhe, Kaiserstr. 12, D-76128 Karlsruhe, Germany}
\vfil
\abstract
In the minimal supersymmetric model (MSSM), the upper limit of the
lightest Higgs boson mass, $\mhl$,
depends strongly on the top quark mass, $m_t$.
We have computed the dominant two-loop radiative corrections to this upper
limit of $\mhl$ in order to eliminate large uncertainties due to QCD
and $m_t^6$ corrections. It is shown that the QCD corrections
significantly reduce
the one-loop corrections. As a result, the SUSY parameter space accessible
to LEP experiments is significantly increased.
\endpage

\REF\susyrev{H.P. Nilles, {\sl
Phys.~Rep.} {\bf 110}, 1 (1984);
H.E. Haber and G.L.
Kane, {\sl Phys.~Rep.} {\bf 117}, 75 (1985);
R. Barbieri, {\sl Riv. Nuovo Cimento} {\bf 11}, 1 (1988) .}
\REF\hhg{J.F. Gunion, H.E. Haber, G.L. Kane and S. Dawson,
{\it The Higgs Hunter's Guide} (Addison-Wesley Publishing Company,
Reading, MA, 1990).}

In recent years
supersymmetric theories have become maybe the most popular
alternatives to the standard model (SM) of elementary particle physics.
In the minimal supersymmetric extension of the SM (MSSM)\refmark\susyrev\
the Higgs sector contains only the two doublets, $H_1$ and $H_2$,
 required to give masses
to up and down type fermions
with all the quartic couplings
related to the
gauge couplings. This leads to various restrictions
among the Higgs masses and couplings\refmark\hhg. The most important
consequence is the existence of a well defined tree-level upper limit
for the mass of the lightest Higgs boson
\REF\prlhh{H.E. Haber and R. Hempfling, {\sl Phys. Rev. Lett.} {\bf 66},
1815 (1991);
Y. Okada, M. Yamaguchi and T. Yanagida,
{\sl Prog. Theor. Phys.} {\bf 85}, 1 (1991);
J. Ellis, G. Ridolfi and F. Zwirner, {\sl Phys. Lett.}
{\bf B257}, 83 (1991).}%
$$
\mhl\leq m\abs{\cos 2\beta}\leq\mz\,,
\qquad\hbox{with }m\equiv \min \{\mz,\mha\} \,.\eqn\treelineq
$$
\REF\mtcorr{H. Arason \etal, \sl Phys. Rev. \bf D46\rm , 3945 (1992).}%
However, it has been shown recently that radiative corrections
can significantly alter this prediction\refmark{\prlhh}.
In particular, the experiments at LEP200 may not
be able to detect $\hl$ or rule out the MSSM.
The region in the SUSY parameter space, than can be ruled out
at LEP experiments
depends crucially on the top quark mass, $m_t$.
However, there is a significant uncertainty in $m_t$ due to large QCD
corrections (\eg\ the running mass differs from the pole mass
by $\calo(7\%)$\refmark\mtcorr).
Eliminating this uncertainty by defining $m_t$ as the pole mass at
the one-loop level requires an explicit two-loop calculation of $\mhl$.
\REF\newcdf{P. Tipton, CDF Collaboration, invited talk given at the
Lepton-Photon-Interaction Conference in Cornell, Ithaca, NY, (1993).}%
\REF\mtconstraint{J. Bagger, S. Dimopoulos and E. Masso, \sl Phys. Lett.
\bf 156B\rm , 357 (1985); \sl Phys. Rev. Lett. \bf 55\rm , 920 (1985).}

The case $\tanb\to\infty$ is particularly
interesting because here the tree-level constraint $\mhl\le\mz$ is
saturated as long as $\mha\gsim\mz$
and thus we expect this case to yield the maximum Higgs mass.
Note that $\mhl$ has another maximum in the limit
$\tanb=0$ (\ie, $v_2=0$).
However, the constraint that the Yukawa couplings do not develop a
Landau-pole at high energies, together with the experimental lower
bound on $m_t>113~\gev$ (95\%\ CL)\refmark\newcdf\
require that
$\tanb\gsim 0.5$\refmark\mtconstraint.
On the other hand,
$\beta\approx\pi/2$ is theoretically
very favorable since it would explain the large ratio of $m_t/m_b$
({\it e.g.} grand unified theories based on SO(10)
predict $\tanb = m_t/m_b$).

\REF\refrgetwo{
J.R. Espinosa and M. Quir\'os,
{\it Phys. Lett.} {\bf B266}, 389 (1991).}%
\REF\refrgeimp{
H.E. Haber, in {\it
Proceedings of the International Workshop on Electroweak Symmetry
Breaking}, Hiroshima, Japan, 12-15 November 1991, edited by
W.A. Bardeen, J. Kodaira and T. Muta (World Scientific,
Singapore, 1992), p.~225;
H.E. Haber and R. Hempfling, {\it Phys. Rev.} {\bf D48}\rm , 4280 (1993).}%
The numerical analysis of the one-loop corrections
\refmark{\prlhh;\refrgetwo;\refrgeimp}\
shows that the dominant contributions
to $\mhl$ come from the top-stop sector due to an $g_t^2 m_t^2$
dependence ($g_t$ is the top Yukawa coupling).
Thus we expect the dominant two-loop corrections
to be the contributions proportional to
$g_{t}^4 m_t^2$ and $g_{t}^2 g_{\rm s}^2 m_t^2$
($g_{\rm s}$ is the QCD gauge coupling).
These terms can be obtained most easily in the approximation where we set
$g$ and $g^{\prime}$ to zero.
In this case the tree-level
potential of the lightest Higgs doublet reduces simply to
$$
V^{(0)} =m_2^2 H_2^\dagger H_2\,.\eqn\redpot
$$
(Note that here $m_2 = \mhl = 0$ at tree-level.)
After shifting the neutral CP-even component by the
vacuum expectation value (VEV) $[H_2\to (h^0+v_0)/\sqrt{2}]$
we obtain
$$\eqalign{
   &V^{(0)}_{h^0}=th^0 +\half
     m_{h^0 0}^2(h^0)^2+\calo[(h^0)^3]\,,\cr}\eqn\vzb
$$
where $m_{\hl 0}^2 = m_2^2$ and $t =  v_0 m_2^2$.
If we eliminate $m_2$ in favor of $t$ we end up with the relation
$m_{h^0 0} = t/v_0$.
Remember that $m_{h^0 0}$
is still an unrenormalized  parameter
which has to be expressed in terms of a physical observable.
This is the physical mass of $\hl$ (without a subscript $0$)
which is identified in the usual
way as the pole of the propagator
\FIG\figinv{The definition of the gauge boson
top quark and scalar self energies, and the Higgs tadpole.
The left and right handed projection operators are
$R,L = (1\pm\gamma_5)/2$.
}%
$$
\mhl^2 = m_{\hl 0}^2 + {\rm Re}~A_{\hl\hl}(\mhl^2)\,.
\eqn\mhp$$
The self-energies and one-point functions (tadpoles) are defined in
fig.~\figinv.
Note that $\hl$ is a stable particle at tree-level
(the decay of $\hl$ into gluons is induced at the one-loop level, and
we can assume that
$\mhl\le 2\min\{m_t,m_{\tilde t_1},m_{\tilde t_2}\}$).
Thus, eq.~\mhp\ is valid even at the two-loop level.
We now demand that $v_0$ is the true VEV to all orders
in perturbation theory.
This means that the tadpoles
corresponding to a Higgs field disappearing into the
vacuum are absent. This reads $t+ A_{h^0}(0)=0$ or
$m_{\hl 0}^2 = m_2^2 = -A_{\hl}(0)/v_0$.
Thus the two-loop renormalized Higgs mass can be written as
$$
\mhl^2 = A_{\hl\hl}(\mhl^2) - {A_{\hl}(0)\over v_0}\,.
\eqn\mhp$$
The calculation of the two-loop diagrams can be simplified considerably
by using the fact that scalar one-particle irreducible
n-point Green functions with
zero external momenta can
be obtained as derivatives from the effective potential, $V_{\rm eff}$,
with respect to the corresponding fields, \eg
$$\eqalign{
& {\d   V_{\rm eff}\over \d   v_0} = m_2^2 v_0 + A_{\hl}(0)
\qquad\hbox{and}\qquad
 {\d^2 V_{\rm eff}\over \d^2 v_0} = m_2^2 +A_{\hl\hl}(0) \,.}\eqn\aeqv
$$
Furthermore, it is easy to show that
$$
\left({\d^2 \over \d v_0^2}- {\d \over v_0\d v_0}\right)V_{\rm eff}
= 4 {m_{t0}^4\over v_0^2}\left({\d \over \d m_{t0}^2}\right)^2 V_{\rm eff}\,,
\eqn\dveqdu
$$
where $V_{\rm eff}$ on the right hand side is
a function of the masses and coupling constants,
but not of $v_0$. Thus we arrive at
$$
\mhl^2 = 4 {m_{t0}^4\over v_0^2}\left({\d\over \d m_{t0}^2}\right)^2
V_{\rm eff}+A_{\hl\hl}(\mhl^2)-A_{\hl\hl}(0)\,.
\eqn\defmhl$$
The advantage of using the effective potential is that
we only have to compute the difference
$A_{\hl\hl}(\mhl^2)-A_{\hl\hl}(0)$.
Note that in a two-loop calculation it is sufficient to
compute only the one-loop contributions to the
Higgs self energies since $\mhl$ here is generated at one-loop.
In particular, we have to keep the $\calo(\epsilon)$ terms
of the one-loop radiatively corrected Higgs mass, $\mhl$,
on the right hand side of eq.~\defmhl,
since it occurs in a divergent expression.
In addition, we have to compute the effective potential
to second order [$V_{\rm eff}=V^{(0)}+V^{(1)}+V^{(2)}$].
The results of this calculation are presented in Appendix~B.
In the terms derived from $V^{(2)}$ we are again
allowed to replace the unrenormalized quantities by the
physical ones.
However, $V^{(1)}$ depends on the unrenormalized top quark mass,
$m_{t0}$, the left- and right-handed top squark masses, $M_{\tilde t_P0}$
($P = L,R$; we ignore the possibility of $L$-$R$ mixing due to trilinear
Higgs-squark-squark coupling, $A_t$, which yields only non-logarithmic and
thus negligible corrections), and the VEV, $v_0$, at the one--loop level.
Thus, we have to express the bare quantities in terms of
physical quantities to first order.
The unrenormalized masses can easily be expressed in terms of their pole
masses.
\REF\pdg{K. Hisaka \etal\ [Particle Data Group], \sl Phys. Rev. \bf
D45\rm , S1 (1992).}%
For the renormalization of the VEV we choose the tree-level relation
$v_0 = 2^{-1/4} G_\mu^{-1/2}$,
where $G_\mu = (1.166\,39\pm0.000\,02)\times10^{-5}$ GeV$^{-2}$\refmark\pdg\
is the fermi-constant
(note that the only contributions to the $\mu$--decay
come through the $W$ self-energy). Therefore, our renormalization conditions
are
$$\eqalign{
&v_0^2\equiv 2^{-1/2}G_\mu^{-1}\left(1-a_{\rm w}\right)\,,\cr
&m_{t0}^2 = m_t^2\left(1 - a\right)\,,\cr
&M_{\tilde t_P 0}^2 = M_{\tilde t_P}^2\left(1-a\ls{P}\right)\,,\qquad
P=L,R\,,
}\eqn\defct$$
where we have introduced the abbreviations
$$\eqalign{
&a\ls{P} = {{\rm Re}\,A_{\tilde t_P \tilde t_P}(\tilde u\ls{P})
\over\tilde u\ls{P}}\,,\quad
a_{\rm w}\equiv {{\rm Re}\,A_{\rm ww}(0)\over \mw^2}\,,\cr
&a \equiv  {\rm Re}\,[\Sigma_{t}^l(u) + \Sigma_{t}^r(u)
-  \Sigma_{t}^L(u) - \Sigma_t^R(u)]\,,
}\eqn\defa$$
and we have defined
$u\equiv m_t^2$ and $\tilde u\ls{P}\equiv M_{\tilde t_P}^2$ ($P = L,R$).
It is now straightforward to derive an expression for the two-loop
radiatively corrected Higgs mass in terms of physical quantities
$$\eqalign{
&\left(\mhl^2\right)_{\rm 2LP} =
   h \left(1 + a_{\rm w} - 2 a \right)\cr
&
+ 4\sqrt{2} N_c u^2 G_\mu\kappa^{-1}
\biggl\{\left(2 a-a\ls{R}-a\ls{L}\right)
\cr
&
+\epsilon\left[(a - \half a_{\rm w}- \half a_h)
\left((\lnbar\tilde u\ls{R})^2+(\lnbar\tilde u\ls{L})^2-2(\lnbar u)^2 \right)
\right.\cr
&\left.
+  a\ls{L}\lnbar \tilde u\ls{L}
+  a\ls{R}\lnbar \tilde u\ls{R}
-2 a      \lnbar u\right] \biggr\}\cr
&+ 4 \sqrt{2} u^2 G_\mu\left({\d\over \d u}\right)^2
V^{(2)}+A_{\hl\hl}(h)-A_{\hl\hl}(0)\,,
}\eqn\hone$$
where $\kappa \equiv 16 \pi^2$ and we have defined
$$
a_h \equiv \left.{\d A_{\hl\hl}(p^2)\over \d p^2}\right|_{p^2 = h}\,.\eqn\defah
$$
We have denoted the one-loop radiatively corrected squared Higgs mass
in the approximation introduced above by
$$
h \equiv 8 \sqrt{2}N_c u^2 G_\mu\kappa^{-1}t_0
\,.
\eqn\abbrev$$
Here we have defined $t_0 \equiv \ln(\msusy^2/m_t^2)$ where
we found it convenient to parameterize our numerical results by
$\msusy^2 \equiv M_{\tilde t_L}M_{\tilde t_R}$
and $r \equiv M_{\tilde t_L}/M_{\tilde t_R}$.
The analytic results for the relevant self-energies and the
expression for the one-loop and
the two-loop effective potential are given in Appendix~A and B,
respectively. Here we will simply present an approximate result in the case
of heavy, mass-degenerate superpartners
$M_{\tilde t_L}^2 = M_{\tilde t_R}^2 = M_{\tilde g}^2 \gg m_t^2$
$$
\eqalign{
&\left(\mhl^2\right)_{\rm APP} =
 h \left\{1- 6 g_{\rm s}^2 C_F \kappa^{-1} \left[ t_0 + 2\right]\right.\cr
&\left.+\sixth g_t^2\kappa^{-1} \left[9 t_0 + 18 + 7 N_c +
 \left(15 - \pi^2 \right)
t_0^{-1}\right]\right\}
\,.}\eqn\approxim
$$
The values of the higgsino mass parameter, $\mu$, and of the heavy Higgs
doublet, $\mhh$, turn out to be irrelevant.

\FIG\figmt{
The two-loop shift in the Higgs mass, $\dmh$, as a function of the top mass.
We present the terms to order
$g_{t}^4 m_t^2$ (Yukawa) and $g_{t}^2 g_{\rm s}^2 m_t^2$ (QCD)
separately.}%
We will now proceed to compare our diagrammatic two-loop result
with the second order terms obtained by a renormalization group
(RG) approach.
Here we reintroduce the Higgs self coupling, $\lambda$, in the potential of
eq.~\redpot\
which we treat as a
running low energy effective parameter rather than a bare parameter.
In this case we find in general that
$\lambda\neq 0$ at any energy
scale $\sqrt{s}<\msusy$ even in the case $g = g^\prime = 0$.
The RG improved Higgs mass can
be written as
$$
\left(\mhl^2\right)_{\rm RG} = \lambda v_0^2\,,\eqn\mhlrge
$$
where the running quartic Higgs self coupling evaluated
at the scale $s = \mhl^2$ is
$$\lambda = \int_{\ln \msusy^2}^{\ln m_t^2} \d t
\beta_{\lambda}\,,\eqn\runningl
$$
\REF\refbeta{%
M.B. Einhorn and D.R.T. Jones, \sl Nucl. Phys. \bf B196\rm, 475 (1982),
and references therein.}
\REF\refbetb{%
C. Ford, I. Jack and D.R.T. Jones,
\sl Nucl. Phys. \bf B387\rm , (1992) 373.}%
By using the $\beta$ function to two-loop order,
$\beta_\lambda$, given in ref.~\refbeta\ and \refbetb\
we obtain the leading and
next-to-leading log radiative corrections to the Higgs mass summed to all
orders in perturbation theory. In order to obtain an analytic result
we will solve the RG equations (RGE) iteratively.
The second order
contributions to order $g_t^6$ and $g_s^2g_t^4$ are
$$
\eqalign{
\left(\mhl^2\right)_{\rm 1\beta} &=
 h \left\{1- t_0 \kappa^{-1} \left[6 g_{\rm s}^2 C_F - \threehalf g_t^2\right]
\right\}\cr
\left(\mhl^2\right)_{\rm 2\beta} &=
\left(\mhl^2\right)_{\rm 1\beta} +
 h \kappa^{-1} \left[4 g_{\rm s}^2 C_F - 5 g_t^2\right]
\,,}\eqn\defrgitwo$$
where the subscript $1\beta$ ($2\beta$) indicates the use of one-loop
(two-loop) $\beta$ functions.
It is instructive to compare eq.~\defrgitwo\ to
eq.~\approxim. We see that the coefficients of $t_0^2$
obtained by solving the one-loop RGEs to second order\refmark\refrgeimp\
are in agreement as they should be.
However, the terms linear in $t_0$
obtained by solving the two-loop RGEs to first order\refmark\refrgetwo\
are different.
This is not surprising since
the RG approach does not include any threshold effects.
For example, the QCD threshold corrections to
$m_t$ are\refmark\mtcorr
$$
m_t = m^{\rm run}_t \left(1+C_F{\alpha_{\rm s}\over \pi}\right)\,,\eqn\runningt
$$
where the running top quark mass is
$m_t^{\rm run} = 2^{-3/4}G^{-1/2}_{\mu} g_t(m_t)$
and we use $\alpha_{\rm s} = g_{\rm s}^2/4\pi = 0.11$.
We now investigate the origin of the $m_t^4$ dependence of the one-loop
corrections to $\mhl$ more closely.
{}From eq.~\mhlrge, \runningl\ and \runningt\ we obtain
$$\eqalign{\mhl^2 &
 \propto g_t^4 G_\mu^{-1} t_0
 \propto m_t^4 G_{\mu} t_0
\left(1-4 C_F{\alpha_{\rm s}\over \pi}\right)
}\,. \eqn\thresh
$$
The $\alpha_{\rm s}$-term in eq.~\thresh\ coming from finite threshold
corrections to $m_t$ are of the same order as the next-to-leading log term
of eq.~\defrgitwo.
If we add both terms together we indeed recover the coefficient proportional
to $g_{\rm s}^2$ in eq.~\approxim.

We will now present the numerical results of our two-loop calculation.
It is contrasted with the second order leading log result and
next-to-leading log result obtained from a simple analysis of the one-loop
RGEs\refmark\refrgeimp\ and
two-loop RGEs\refmark\refrgetwo\ not including QCD threshold
corrections [eq.~\defrgitwo].
We set $\mu = \mhh = 200~\gev$, $r = 1$ and
$M_{\tilde b_L}^2 = M_{\tilde g}^2 = \msusy^2 - m_t^2$.
In fig.~\figmt\ we present the two-loop contribution to the Higgs mass
obtained by different methods.
Presented are the results from our diagrammatic two-loop calculation (solid
curve), the second order contributions from a RG approach using one-loop
and two-loop $\beta$~functions (dotted curve and dot-dashed curve,
respectively) and result of our approximate expansion (dashed curve).
We have parameterized the results by a linear mass shift defined as
$$
\left(\Delta\mhl\right)\ls{x} = {(\mhl^2)\ls{x} - h\over 2 \mz}\,,
\quad\hbox{ $x$ = 2LP, $1\beta$, $2\beta$, APP}
\,.\eqn\linear
$$
The contributions to order $g_{t}^4 m_t^2$ (Yukawa)
and $g_{t}^2 g_{\rm s}^2 m_t^2$ (QCD) are plotted separately.
The dependence of $\Delta\mhl$
on the remaining parameters, $r$,
$M_{\tilde g}$, $\mu$ and $\mhh$
changes the result by less than 1 GeV as long as $M_{\tilde g},\mu <
3 \msusy$ and  $0.1<r<10$.

\REF\sher{M. Sher, \sl Phys. Lett. \bf B140\rm , 334 (1984).}%
Sofar we have focused our attention on the case of large $\tanb$
where one obtains the upper limit for $\mhl$.
We have seen that the diagrammatic result differs significantly from the
RGE results [eq.~\defrgitwo]. However,
the agreement with our large $\msusy$ expansion [eq.~\approxim]
is excellent.
Furthermore, the two-loop result is
dominated by the QCD corrections which are even larger than those one-loop
electroweak corrections which are not enhanced by powers of $m_t$
and thus never exceed a few GeV\refmark\sher.

\FIG\figmsusy{
The radiatively corrected Higgs mass,
$\mhl$, as a function of $\msusy$.
Included are terms to order
$g_{t}^4 m_t^2$ and $g_{t}^2 g_{\rm s}^2 m_t^2$.}%
In recent years it has become standard in
experimental analyses to only include the dominant $m_t^4$ term
at the leading log level via the CP-even mass matrix in electroweak
eigenstates
$$\eqalign{
{\cal M}^2 &=
 m_{A^0}^2\left(\matrix{\sin^2\beta&-\sinb\cosb\cr
-\sinb\cosb&\cos^2\beta}\right)
\cr &
+\mz^2\left(\matrix{\cos^2\beta &-\sinb\cosb\cr
-\sinb\cosb&\sin^2\beta +\epsilon \mz^{-2}}\right)
\,,}
\eqn\massmhh
$$
where $\epsilon = h/\sin^2\beta$.
Note that at this level our two-loop calculation can be directly
generalized to the case of arbitrary $\tanb$ by using
$$
\epsilon
 \equiv {3 \sqrt{2}\over 2 \pi^2} {m_t^4\over \sin^2\beta} G_\mu t_0
\left[1 + {2 \alpha_{\rm s}\over \pi}\left(t_0 + 2\right)\right]^{-1}
\,,\eqn\defeps
$$
\REF\hhh{H.E. Haber, R. Hempfling and A.H. Hoang, work in progress.}%
where
$t_0 \equiv \ln (M_{\tilde t_1} M_{\tilde t_2} / m_t^2)$.
In fig.~\figmsusy\ we compare this approximation (solid curve)
with the one-loop result, $\sqrt{\mz^2+h}$, (dotted curve)
and the two-loop result, $\sqrt{\mz^2+(\mhl^2)_{\rm 2LP}}$, (dashed curve)
in the limit of large $\tanb$.
We plot $\mhl$ as a function of $\msusy$ for $m_t = 150~\gev$ and $r = 1$,
and we set $\mhh^2 = \mu^2 =
M_{\tilde b_L}^2 = M_{\tilde g}^2 = \msusy^2 - m_t^2$.
We see that in the few TeV region the two-loop corrections are comparable
to the one-loop result in magnitude due to large logarithms.
Those terms have to be summed over using the RGEs.
We have moved the QCD corrections into the denominator which is in better
agreement with the RG improved result\refmark\hhh.
In particular, the approximation of eq.~\massmhh\ and \defeps\ increases
monotonically even for very large $t_0$.

\FIG\figcont{
Contours of constant Higgs mass, $\mhl$, in the
$\tanb$--$\msusy$ plane.
We take $m_t = 150~\gev$ and consider the limit $\mha^2 \gg \mz^2$.}%
In fig.~\figcont\ we
show contours of constant $\mhl =60$, 80, 100, 110, 120, 130~GeV
in the $\tanb$--$\msusy$ plane
in the approximation from eq.~\massmhh.
We compare the results including QCD corrections
(solid curves) and without QCD corrections (dashed curves)
in the limit $\mha^2 \gg \mz^2$ where we obtain the upper limit of $\mhl$.

We have shown that the QCD corrections to $\mhl$
are large and negative and thus lead to
a reduction of the one-loop result by $\calo(30 \%)$ for $\msusy = 1~\tev$.
Consequently the region in the SUSY parameter space accessible at LEP
experiments increases significantly. The effects are particularly
important in the case $\tanb = 1$ where the tree-level Higgs mass vanishes.
The implications for the large $\tanb$ regime depend
on whether it is possible to detect a Higgs boson with a mass,
$\mhl<\mz$.
Consider \eg\ the scenario where $m_t = 150~\gev$
and the experimental lower limit
of $\mhl\geq 110~\gev$. Then the QCD corrections shift the lower limit
of $\msusy$ form 670~\gev to 1.3~\tev.

\ack{We would like to thank H. K\"uhn for suggesting this project.}

\Appendix{A}

\REF\reduct{W. Siegel, \sl Phys. Lett. \bf B84\rm, (1979) 193;
D.M. Capper, D.R.T. Jones and P. van Nieuwenhuizen,
\sl Nucl. Phys. \bf B167\rm , (1980) 479.}%
In this appendix we present the results of the
one-loop self energies of top quark, top squarks, the Higgs boson
and $W$ boson
in the approximation that $g = g^{\prime}= A_U = 0$.
We have done all the calculations in dimensional reduction
in  order to preserve SUSY (\ie, we set the number of space-time dimension to
$n= 4-2\epsilon$, but keep the number of components of all other
tensors fixed)\refmark\reduct.
We have suppressed the
arbitrary renormalization scale.
The results are presented in terms of the
following integrals formulated in Minkowski space using the Bjorken-Drell
metric
$$\eqalignno{
&A_0(m^2) =-i \kappa\int{\d^n k\over
(2\pi)^n}{1\over k^2-m^2+i\delta}\,,\cr
&B_0(q^2,m_1^2,m_2^2)=-i\kappa \int{\d^n k\over (2\pi)^n}
{1\over (k^2-m_1^2+i\delta)[(k+q)^2-m_2^2+i\delta]}\,,\cr
&q_{\mu}B_{1}(q^2,m_1^2,m_2^2)
=-i\kappa \int{\d^n k\over (2\pi)^n}{k_{\mu}\over
(k^2-m_1^2+i\delta)[(q+k)^2-m_2^2+i\delta]}
\,.&\eq}
$$

The top quark self energies are
$$\eqalign{
&\kappa\Sigma_{t}^{\rm L}(u) =
-g_{t}^2\left[B_1(u,\tilde h,\tilde u\ls{R})+B_1(u,u,0)\right]\cr
&
 + g_{\rm s}^2 C_F
\left[-2 B_1(u,\tilde g,\tilde u\ls{L})
\right.\cr
&\left.
+(1+\xi)B_0(u,0,u)+2 B_1(u,0,u)\right]
\,,\cr}
\eqn\selftt$$
$$\eqalign{
&\kappa\Sigma_{t}^{\rm R}(u) =
-g_{t}^2\left[B_1(u,\tilde h,\tilde u\ls{L})
+B_1(u,\tilde h,\tilde d\ls{L})
\right.\cr
&\left.
+B_1(u,u,0)+B_1(u,0,0)\right]
\cr
&
 + g_{\rm s}^2 C_F
\left[-2 B_1(u,\tilde g,\tilde u\ls{R})
\right.\cr
&\left.
+(1+\xi)B_0(u,0,u)+2 B_1(u,0,u)\right]
\,,
}\eqn\selft$$
$$\eqalign{
&\kappa\Sigma_{t}^{\rm l}(u) = g_{\rm s}^2 C_F
(3+\xi)B_0(u,0,u)\,,
}\eqn\slt$$
$$\eqalign{
&\kappa\Sigma_{t}^{\rm r}(u) = g_{\rm s}^2 C_F
(3+\xi)B_0(u,0,u)\,.
}\eqn\srt$$
The top squark self energies are
$$\eqalignno{
&\kappa A_{\tilde t_L\tilde t_L}(\tilde u\ls{L})=
 g_{\rm s}^2 C_F
\left\{-A_0(\tilde u\ls{L})\right.\cr
&\left.
+\left[4 u_{\tilde L}
B_0(\tilde u\ls{L},0,\tilde u\ls{L})-A_0(\tilde u\ls{L})\right]
\right.\cr
&\left.
+2\left[\left(u+\tilde g-\tilde u\ls{L}\right)
B_0(\tilde u\ls{L},\tilde g,u)+A_0(\tilde g)+A_0(u)\right]\right\}\cr
&+g_{t}^2\left\{
\left[A_0(u)+A_0(\tilde h)
+(u+\tilde h-\tilde u\ls{L})B_0(\tilde u\ls{L},\tilde h,u)\right]
\right.\cr
&\left.
-2  u B_0(\tilde u\ls{L},\tilde u\ls{L},0)
-   u B_0(\tilde u\ls{L},\tilde d\ls{L},0)
-\tilde h B_0(\tilde u\ls{L},\tilde u\ls{R},H)
\right.\cr
&\left.
- A_0(\tilde u\ls{R})
\right\}\,,
&\eq}$$
$$\eqalignno{
&\kappa A_{\tilde t_R\tilde t_R}(\tilde u\ls{R})=
 g_{\rm s}^2 C_F
\left\{-A_0(\tilde u\ls{R})\right.\cr
&\left.
+\left[4 u_{\tilde R}
B_0(\tilde u\ls{R},0,\tilde u\ls{R})-A_0(\tilde u\ls{R})\right]
\right.\cr
&\left.
+2\left[\left(u+\tilde g-\tilde u\ls{R}\right)
B_0(\tilde u\ls{R},\tilde g,u)+A_0(\tilde g)+A_0(u)\right]\right\}\cr
&+g_{t}^2\left\{
\left[A_0(u)+A_0(\tilde h)
+(u+\tilde h-\tilde u\ls{R})B_0(\tilde u\ls{R},\tilde h,u)\right]
\right.\cr
&\left.
+\left[A_0(\tilde h)
+(\tilde h-\tilde u\ls{R})B_0(\tilde u\ls{R},\tilde h,0)\right]
\right.\cr
&\left.
-2  u B_0(\tilde u\ls{R},\tilde u\ls{R},0)
-  \tilde h B_0(\tilde u\ls{R},\tilde d\ls{L},H)
-  \tilde h B_0(\tilde u\ls{R},\tilde u\ls{L},H)
\right.\cr
&\left.
- A_0(\tilde u\ls{L})
- A_0(\tilde d\ls{L})
\right\}\,.
&\eq}$$
The Higgs boson self energy is
$$
\eqalign{
&\kappa A_{\hl\hl}(h) = N_c g_{t}^2
\left\{\left[
\left(4 u-h\right)B_0(h,u,u)+2 A_0(u)\right]
\right.\cr
&\left.
-2u B_0(h,\tilde u\ls{L},\tilde u\ls{L})
-2u B_0(h,\tilde u\ls{R},\tilde u\ls{R})
\right.\cr
&\left.
-A_0(\tilde u\ls{L})-A_0(\tilde u\ls{R})\right\}\,.
}\eqn\ahh$$
The $W$ self energy is
$$\eqalign{
\kappa A_{WW}(0) = N_c \mw^2 g_{t}^2
\left({1\over\epsilon}+\half
 - \lnbar u + {\tilde u\ls {L}+\tilde d\ls{L} \over 2 u}
-{\tilde u\ls{L}\tilde d\ls{L} \over u^2}
\ln{\tilde u\ls{L} \over \tilde d\ls{L}}\right)\,,}
\eqn\aww$$
where we abbreviate
the gluino mass, the left-handed bottom squark,
the Higgsino mass and the mass of the second Higgs doublet mass by
$\tilde g\equiv M_{\tilde g}^2$,
$\tilde d\ls{L}\equiv M_{\tilde b_L}^2$,
$\tilde h \equiv \mu^2$ and
$H     \equiv \mhh^2$, respectively.

\Appendix{B}

Here we present the calculation of the Feynman diagrams
contributing to the effective potential. We
have done all the calculations in dimensional reduction to preserve
the supersymmetric Ward-identities.
The QCD gauge sector has been calculated in arbitrary
$R_\xi$ gauge. The results
for the one-loop and two-loop effective potential are presented in terms of the
following integrals
$$\eqalign{
&K(m^2)= i \int{\d^{n}k\over
(2\pi)^{n}}\ln(k^2-m^2+i\delta)= \int_0^{m^2}{\d x\over \kappa} A_0(x)\,,
}\eqn\defonelp$$
and \refmark\refbetb
$$\eqalignno{
&\kappa^2 J(m^2_1,m_2^2) = A_0(m_1^2) A_0(m_2^2)\,,\cr
&\kappa^2 I(m^2_1,m_2^2,m_3^2)=\kappa^2 \int {\d^{n}k\over (2\pi)^{n}}
{\d^{n}p\over (2\pi)^{n}}
(k^2-m_1^2+i\delta)^{-1}
(p^2-m_2^2+i\delta)^{-1}
[(k+p)^2-m_3^2+i\delta]^{-1}\,.\cr &
&\eq}$$
Furthermore, we find it convenient to introduce
$$
L(a,b,c) \equiv J(c,b)-J(a,b)-J(a,c)-(a-b-c) I(a,b,c)\,.
\eqn\defl$$
The one-loop effective potential is
$$
V^{(1)} = N_c\left[2 K(u)
- K(\tilde u\ls{L})
- K(\tilde u\ls{R})\right]\,,
\eqn\vef$$
The relevant QCD contributions
to the two-loop effective potential are
$$\eqalignno{
&V^{(2)}_{\rm s} =  g_{\rm s}^2(N_c^2-1)\left\{
\left[-2u I(u,u,0)+ J(u,u)\right]
\right.\cr
&\left.
+\left[\tilde u\ls{L} I(\tilde u\ls{L},\tilde u\ls{L},0)
+\fourth J(\tilde u\ls{L},\tilde u\ls{L})\right.
\right.\cr
&\left.
+\left.\tilde u\ls{R} I(\tilde u\ls{R},\tilde u\ls{R},0)
+\fourth J(\tilde u\ls{R},\tilde u\ls{R})\right]
\right.\cr
&\left.
+\left[L(\tilde u\ls{R},\tilde g,u) + L(\tilde u\ls{L},\tilde g,u)\right]
\right.\cr
&\left.
+\fourth\left[
J(\tilde u\ls{L},\tilde u\ls{L})+J(\tilde u\ls{R},\tilde u\ls{R})
\right]\right\}\,,
&\eq}
$$
and the second order top Yukawa contributions are
$$
\eqalignno{
&V^{(2)}_{t} = N_c g_{t}^2\left\{
\left[L(0,u,u)+L(0,0,u)
     +L(\tilde u\ls{L},\tilde h,u)
\right.\right.\cr
&\left.\left.
     + L(\tilde u\ls{R},\tilde h,u)
     + L(\tilde d\ls{L},\tilde h,u)
     + L(\tilde u\ls{R},\tilde h,0)\right]
\right.\cr
&\left.
- \left[ u I(\tilde u\ls{L},\tilde u\ls{L},0)
       + u I(\tilde d\ls{L},\tilde u\ls{L},0)
       + u I(\tilde u\ls{R},\tilde u\ls{R},0)
\right.\right.\cr
&\left.\left.
       + \tilde h I(\tilde u\ls{L},\tilde u\ls{R},H)
       + \tilde h I(\tilde d\ls{L},\tilde u\ls{R},H)\right]
\right.\cr
&\left.
+\left[J(\tilde u\ls{L},\tilde u\ls{R})
      +J(\tilde d\ls{L},\tilde u\ls{R})
\right]\right\}\,.
&\eq}
$$
The derivative of $V^{(2)} = V^{(2)}_{\rm s} + V^{(2)}_{t}$
in eq.~\hone\ is obtained by using the chain rule and
$$
{\d \tilde u_L\over \d u} = {\d \tilde u_R\over \d u} =  1\,,
\qquad\hbox{and}\qquad
{\d \tilde d_L\over \d u} = {\d \tilde g\over \d u} =  0\,.\eqn\blabla
$$
We have computed the second derivative of $V^{(2)}$
numerically and analytically. Furthermore, we have
checked both analytically and numerically that the right hand side of
eq.~\hone\ is independent of $\epsilon$ and the renormalization scale.

\noindent{$\underline{\rm Remark:}$}
We have done the two-loop calculation in dimensional reduction.
Operationally this was done using dimensional regularization
plus the contributions from the so-called
$\epsilon$-scalars\refmark\reduct. The latter contribution is divergent due to
the $1/\epsilon^2$ pole from the two-loop integral.
Without it eq.~\hone\ is not finite.
\refout
\figout
\end